\documentclass[twocolumn,secnumarabic,amssymb, nobibnotes, aps, prd]{revtex4-1}
\pdfoutput=1
\usepackage{amsmath}
\usepackage{nicefrac}
\usepackage{graphicx}
\usepackage{wrapfig}
\usepackage{setspace}
\usepackage{subcaption}
\usepackage{url}
\usepackage{csquotes}
\usepackage{amssymb}
\usepackage{float}
\usepackage{titlesec}
\usepackage{indentfirst}
\usepackage{mathrsfs}

\titlespacing*{\section}{0pt}{1.1\baselineskip}{\baselineskip}

\begin{document}

\title{Configurational Entropic Study of the Enhanced Longevity in Resonant Oscillons}

\author{Marcelo Gleiser}
\email[Author's Email: ]{mgleiser@dartmouth.edu}
\affiliation{Department of Physics and Astronomy\\ Dartmouth College, Hanover, NH 03755, USA}

\author{Max Krackow}
\email[Author's Email: ]{Max.E.Krackow.GR@dartmouth.edu}
\affiliation{Department of Physics and Astronomy\\ Dartmouth College, Hanover, NH 03755, USA}

\date{\today}

\begin{abstract}
We use an information-theoretic measure of shape complexity known as configurational entropy (CE) to investigate numerically the remarkably long lifetimes of spherically-symmetric ``resonant oscillons'' in three-dimensional and of azimuthally-symmetric oscillons  in two-dimensional relativistic scalar field theories, which have been conjectured to be infinite. In 3d, we obtain a power law relating a stability measure derived from CE to the oscillons' lifetimes that, upon extrapolation to large times, offers support to this conjecture. In 2d, we obtain a three-way relation between the oscillons' energies, a CE-derived measure of their stability, and their radiation rates to support the conjecture that they asymptotically tend toward a classically-stable attractor solution.
   
\end{abstract}

\maketitle

\section{Introduction}

The study of long-lived, solitonic configurations in field theories has a long history, dating back to Edmond Bour's study of surfaces of constant negative curvature in the mid nineteenth century \cite{Bour}, which was rediscovered in 1939 as the Frenkel-Kontorova model encapsulated in the Gauss-Codazzi equation \cite{Frenkel}. Solitons experienced a resurgence in the 1970s with the possibility that they could model particles with spatial extension \cite{Dashen,Coleman,Rajamaran}. Typical applications involve finding a static solution to a nonlinear field equation describing a specific physical system with its stability determined by the topology of the vacuum manifold.

Above one spatial dimension, no time-independent solution involving only real scalar fields exist, a result known as Derrick's theorem \cite{Derrick,Coleman}. One way to circumvent this limitation is to consider models featuring time-dependent complex scalar fields, coupled or not to other fields. The solutions, known as nontopological solitons, owe their (classical) stability to a conserved global charge $Q$, such as particle number. Examples of the vast literature on nontopological solitons and the related $Q$-balls can be found in Refs. \cite{TDLee,Coleman2}.

Another possibility is to find spatially-bound real scalar field configurations that are time-dependent but still extremely long-lived. This is the case of oscillons, first discovered by  Bogolyubskii and Makhan'kov in 1976 \cite{Bogolyu}, and rediscovered by Gleiser in 1994, where the name ``oscillon'' was first suggested \cite{Gleiser94}. Since then, oscillons have attracted a huge amount of interest due to their potential applications in high-energy physics and cosmology, as this incomplete list of references shows \cite{Copeland, Sornborger, Fodor, Honda, Kasuya, Graham, Sicilia, Farhi, Hertzberg, Salmi, Saffin, Mukaida, Howell, Thorarinson, Amin1, Amin2, Amin3, Amin4, GleiserGrahamStama1}.

\begin{figure}[t]  
   \centering
   \includegraphics[width=0.5\textwidth]{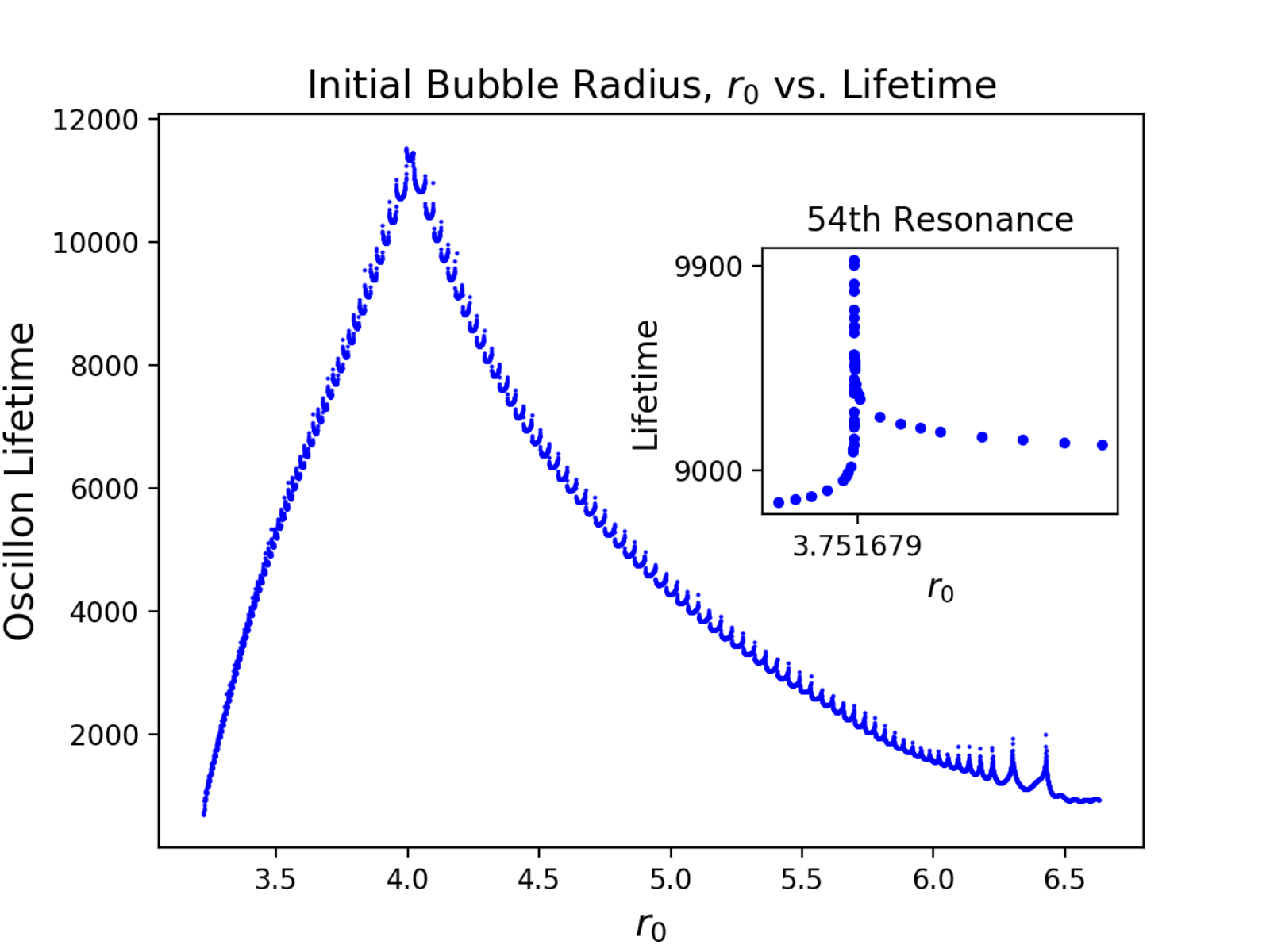}
\caption{The 127 resonances seen along the lifetime vs. $r_0$ plot for 3d oscillons obtained with the potential of Eq. \ref{potential}. The 64th resonance is the highest, or peak, resonance.  The inset shows resonant oscillons generated from many initial radii straddling the 54th resonance.}
\label{resmountain}
\end{figure}

Of the many interesting properties of the oscillons studied so far in the literature, one of the most intriguing is the conjecture, first raised by Honda and Choptuik in Ref. \cite{Honda} (henceforth HC), that a class of three-dimensional, spherically-symmetric scalar-field oscillons with simple double-well potential interactions could be infinitely long-lived. Those oscillons are obtained from generic Gaussian initial configurations with radial parameter $r_0$, and appear as resonant peaks in the oscillon lifetime versus initial radius curve, as shown in Figure \ref{resmountain}. Henceforth, we refer to the lifetime profile depicted in Figure \ref{resmountain} as the ``resonance mountain,'' and to oscillons living in each of the resonant peaks as ``resonant oscillons'' \cite{GleiserKrackow}.

In analogy with the critical point of continuous phase transitions, where the correlation length diverges at the critical temperature $T_c$ \cite{Landau}, HC proposed that, as each resonance is approached from below and from above, there will be a critical value for the initial radius, $r_{0n}^*$, where the lifetime diverges to arbitrarily large values.  Here, $1\leq n\leq 127$, since we observe 127 resonances in total.  As in the numerical study of phase transitions, the approach to the critical values $r_{0n}^*$ is limited by simulations on a field lattice and can only be inferred indirectly. Still, the possibility of infinitely long-lived, self-supporting scalar field configurations is so remarkable and counter-intuitive that it deserves further study.

In a recent work \cite{GleiserKrackow}, we presented supporting evidence for the HC conjecture using parametric resonance and virial methods. We also added a longevity study of two-dimensional oscillons, since these too have yet to be shown to decay in numerical simulations. Motivated by these results, in the present work we extend our studies of the longevity of resonant and 2d oscillons using a recently proposed measure of spatial complexity known as Configurational Entropy (CE) \cite{GleiserStamatopoulos1}, which has been applied to many physical systems, from solitons in field theories \cite{CorreaGleiser,Correa,Bernardini,Braga,Lee, GleiserStephens} to phase transitions \cite{GleiserStamatopoulos2, GleiserSowinski2,GleiserSowinski3}, and to astrophysics and cosmology \cite{GleiserSowinski1,GleiserJiang1,GleiserGraham,BernardiniRocha,KouTianZhou}.

In this work, we use that CE has been successfully used as a phenomenological indicator of stability of localized field configurations in several physical systems:  in the application of CE to stellar polytropes, it was found that as the binding energy of the Newtonian stars decreased, CE increased.  Following this stability trend, CE was used to determine the Chandrasekhar limit for white dwarfs to few percent accuracy \cite{GleiserSowinski1}.  For general-relativistic neutron and boson stars, the critical points of CE nearly paralleled their critical stability regions determined by perturbation theory \cite{GleiserJiang1}.  Additionally, CE has been employed as a predictor of stability in the study of decay rates in hydrogen  \cite{GleiserJiang2}. More to the point, CE was recently shown to be a predictor of lifetimes for non-resonant oscillons \cite{GleiserStephens}. Here, we extend the CE formalism to investigate the HC conjecture for resonant oscillons, which require a different treatment. In Section 2 we describe the model and its numerical implementation. In Section 3, we briefly review configurational entropy. In Section 4 we present our results for three and two-dimensional oscillons. In 3d, we obtain a power-law relation between the resonant oscillon lifetimes and their related CE stability measure, offering further support for the HC conjecture. In 2d, we use CE to show that oscillons radiate their initial energy toward an attractor solution which, at least numerically, is classically stable, as anticipated analytically in Ref. \cite{Sicilia}. We summarize our results in Section 5.

\section{The Model and Its Numerical Implementation}

We work with metric signature (-,+,+,+) and use spherical coordinates. More details of our model and numerical implementation can be found in Ref. \cite{GleiserKrackow}.

The action for a self-interacting, real scalar field in $D = d + 1$ spacetime dimensions is
\begin{equation}
S[\phi] = \int{d^Dx\sqrt{\left|g\right|}\left(-\frac{1}{2}g^{\mu\nu}\partial_{\mu}\phi\partial_{\nu}\phi - V(\phi)\right)},
\end{equation}
\noindent
where we consider the double-well potential,
\begin{equation}
V(\phi) = \frac{\lambda}{4}\phi^2\left(\phi - \phi_0\right)^2.
\end{equation}
Introducing the following dimensionless field and coordinates,
$\phi = \alpha\bar{\phi},~
t = \xi\bar{t},~{\rm and}~
r = \xi\bar{r}$, and defining
$v = \frac{\phi_0}{\alpha}$, $\xi^{-1} = \sqrt{\lambda} \alpha$ and droping the bars, the action becomes,
\begin{equation}
S[\phi] = \lambda^{\frac{1 - d}{2}}\phi_0^{3-d}v^{d-3}\int{d^Dx\sqrt{\left|g\right|}\left(-\frac{1}{2}g^{\mu\nu}\partial_{\mu}\phi\partial_{\nu}\phi - V(\phi)\right)}.
\end{equation}
The physical units of time and distance are then $\xi = v/\sqrt{\lambda}\phi_0$.  Choosing $v = \sqrt{2}$, the dimensionless potential is
\begin{equation}
V(\phi) = \frac{1}{2}\phi^2 - \frac{1}{\sqrt{2}}\phi^3 + \frac{1}{4}\phi^4,
\label{potential}
\end{equation}
with degenerate minima at $\phi = 0$ and $\phi = \sqrt{2}$.

Following HC \cite{Honda}, we transform to monotonically increasing boosted (MIB) coordinates.  
We take
$\tilde{t} = t,~
\tilde{r} = r + f(r)t,~{\rm and}~
\tilde{\Omega} = \Omega$, where $f(r)$ is a smooth, monotonically increasing function interpolating between 0 and 1 at a cutoff radius $r_c$:  $f(r) \rightarrow 0$ when $r \ll r_c$ and $f(r) \rightarrow 1$ for $r \gg r_c$.  Choosing $f(r) = \frac{1}{2}\text{tanh}\left(\frac{r-R}{\delta}\right) - \frac{1}{2}\text{tanh}\left(\frac{-R}{\delta}\right),$ we see that MIB coordinates transition from regular spherical coordinates to light cone coordinates at a distance $R$ from the origin in a space of thickness $\delta$.  The transition to light cone coordinates blue shifts radiation and causes it to take a longer time to bounce back, effectively freezing the radiation in the region of thickness $\delta$.  

Using the definitions
\begin{align}
a(t,r) &= 1 + f'(r)t, \\
\beta(t,r) &= \frac{f(r)}{1 + f'(r)t},
\end{align}
the Euler-Lagrange equation becomes,
\begin{equation}
\begin{split}
& \frac{1}{\sqrt{\left|g\right|}}\partial_{\mu}\sqrt{\left|g\right|}g^{\mu\nu}\partial_{\nu}\phi = \frac{1}{a\tilde{r}^{d-1}}\left(\partial_t\left[a\tilde{r}^{d-1}\left(-\partial_t\phi + \beta\partial_r\phi\right)\right]\right) \\
& + \frac{1}{a\tilde{r}^{d-1}}\left(\partial_r\left[a\tilde{r}^{d-1}\left(\beta\partial_t\phi + \frac{1 - a^2\beta^2}{a^2}\partial_r\phi\right)\right]\right).
\end{split}
\end{equation}
Defining two auxiliary fields, $\Phi$ and $\Pi$ as
\begin{align}
\Phi &= \partial_r\phi, \label{auxiliaryfields1}\\
\Pi &= a\left(\partial_t\phi - \beta\partial_r\phi\right),
\label{auxiliaryfields2}
\end{align}
\noindent
we obtain three coupled equations of motion for $\phi$, $\Phi$, and $\Pi$.  Inserting (\ref{auxiliaryfields1}) into (\ref{auxiliaryfields2}) and rearranging, we arrive at the equation of motion for $\phi$ in terms of the auxiliary fields,
\begin{equation}
\partial_t\phi = a^{-1}\Pi + \beta\Phi.
\label{eom1}
\end{equation}
The two equations for the auxiliary fields are,
\begin{align}
\begin{split}
\partial_t\Phi &= \partial_r\left[a^{-1}\Pi + \beta\Phi\right],\\
\partial_t\Pi &= \frac{1}{\tilde{r}^{d-1}}\partial_r\left[\tilde{r}^{d-1}\left(\beta\Pi + a^{-1}\Phi\right)\right] - \left(d - 1\right)\frac{f}{\tilde{r}}\Pi - a\frac{dV}{d\phi}.
\end{split}
\label{eom2}
\end{align}
Using the auxiliary fields in the definition of the stress energy-momentum tensor, we derive the energy density,
\begin{equation}
\rho = T^{00} = \frac{\Pi^2 + \Phi^2}{2a^2} + V(\phi),
\end{equation}
\noindent
with the gradient energy density and kinetic energy density terms
\begin{equation}
{\cal E}_{\nabla} = \frac{\Pi^2 - \Phi^2}{2a^2},~~{\cal E}_K =
\partial^{0}\phi\partial^{0}\phi = \frac{\Pi^2}{a^2}.
\end{equation}
\noindent
To find resonant oscillons, we evolve $\phi$ and the two auxiliary fields, $\Phi$ and $\Pi$.  Our initial condition is a Gaussian profile interpolating between the two vacua, $\phi(r,t = 0) = \sqrt{2}e^{-\left(r/r_0\right)^2}$. Since oscillons are attractors in field configuration space \cite{Sicilia}, different initial profiles with proper boundary conditions would also generate them. Spatial derivatives are computed using a finite difference method with second order accuracy.  For the time progression, we implement an iterative method for a field f, resulting in a second order scheme (see Ref. \cite{GleiserKrackow})
\begin{equation}
\text{f}^{\, t+1} = \text{f}_{(0)}^{\, t} + \left(\frac{\Delta t}{2}\right)\text{F}\left[\text{f}^{\, t + 1}\right] + \mu_{diss}\left[\text{f}^{\, t}\right].
\end{equation}
The dissipative term, $\mu_{diss}$, takes the form $-\varepsilon\nabla^4\phi\Delta x^3$.  We used $\varepsilon = 0.2$. In $k$-space, we have $\nabla^4\phi \rightarrow k^4\phi_k$: the dissipative term becomes significant at higher $k$ modes suppressing the blue shifted frequencies.

\section{Configurational Entropy in Brief}

In this section we briefly review the mathematical formalism of CE \cite{GleiserStamatopoulos1}. For details see Ref. \cite{GleiserStephens}.  Since we wish to study spatially-localized configurations such as a field or its energy density, consider the set of square-integrable bounded functions $f(\textbf{x}) \in L^2(\textbf{R})$.  Denoting $F(\textbf{k})$ as the Fourier transform of $f(\textbf{x})$, define the modal fraction $f(\textbf{k})$, as
\begin{equation}
f(\textbf{k}) = \frac{\left|F(\textbf{k})\right|^2}{\int\left|F(\textbf{k})\right|^2d^d\textbf{k}}.
\end{equation}
\noindent
For periodic functions with an associated Fourier series, the modal fraction is
$f(\textbf{k}) \rightarrow f_n = \frac{\left|A_n\right|^2}{\sum\left|A_n\right|^2}$,
where $A_n$ is the coefficient of the $n$-th Fourier mode.  In analogy with Shannon's information entropy, the discrete configurational entropy, $S_c[f]$, is defined as \cite{GleiserStamatopoulos1}
\begin{equation}
S_c[f] = -\sum f_n \text{ln}(f_n).
\label{discreteCE}
\end{equation}
\noindent
CE quantifies the spatial informational content of a function modeling a physical system.  It is maximal for a system with $N$ modes carrying the same weight, $f_n = 1/N$,  $S_c = \text{ln}N$. If only one mode $n^*$ carries all the weight, $f_{n^*} = 1$, and CE is minimized, $S_c = 0$.  Lower $S_c$ corresponds to less localization in space.
\par
For continuous systems, the mathematically consistent form of CE is the Differential CE (DCE) \cite{GleiserStephens}, which ensures positive-definiteness. Consider the energy density of a field $\rho (\textbf{r},t)$ in $d$-spatial dimensions with a bounded $l^2$-norm and its Fourier transform, $
\tilde{\rho}(\textbf{k},t) = \left(2\pi\right)^{-\frac{d}{2}}\int d^d\textbf{r} \, \rho (\textbf{r},t) e^{-i\textbf{k} \cdot \textbf{r}}.$
The probability of a wave mode being detected in a volume $d^d\textbf{k}$ centered at $\textbf{k}_+$ is proportional to the power carried by that mode, $p\left(\textbf{k}_+|d^d\textbf{k}\right) \propto \left|\tilde{\rho} (\textbf{k}_+,t)\right|^2 d^d\textbf{k}.$ 
The weight of a mode $\textbf{k}_+$ relative to the mode about which the power spectrum peaks, $\textbf{k}_*$, is defined as the normalized modal fraction,
\begin{equation}
f_{\rm N}(\textbf{k},t) = \frac{ p \left( \textbf{k}_+ | d^d \textbf{k} \right) }{ p \left( \textbf{k}_* | d^d \textbf{k} \right) } = \frac{ \left| \tilde{\rho} ( \textbf{k}_+,t ) \right|^2 }{ \left| \tilde{\rho}  ( \textbf{k}_*,t ) \right|^2 },
\end{equation}
\noindent
bounded as $0 \leq f_{\rm N}(\textbf{k},t) \leq 1$.  We thus define DCE as
\begin{equation}
\begin{split}
C[\tilde{\rho} (\textbf{k},t)] & = - \int d^d \textbf{k} \, f_{\rm N}( \textbf{k},t ) \, \text{ln} f_{\rm N}( \textbf{k},t ) \\
& = C[t].
\end{split}
\end{equation}
\par
In Ref. \cite{GleiserStephens}, different quantities related to DCE were proposed to obtain the general result that the lifetime of oscillons is inversely related to the magnitude of such measures: more stable configurations display lower values of CE.  Due to the essential dynamical differences between normal oscillons and the resonant oscillons we study here, we must approach the problem differently, as explained in what follows.

\section{Lifetime of Resonant Oscillons}

\noindent

\subsection{3d DCE Analysis}

\begin{figure}[H]
   \centering
   \includegraphics[width=0.5\textwidth]{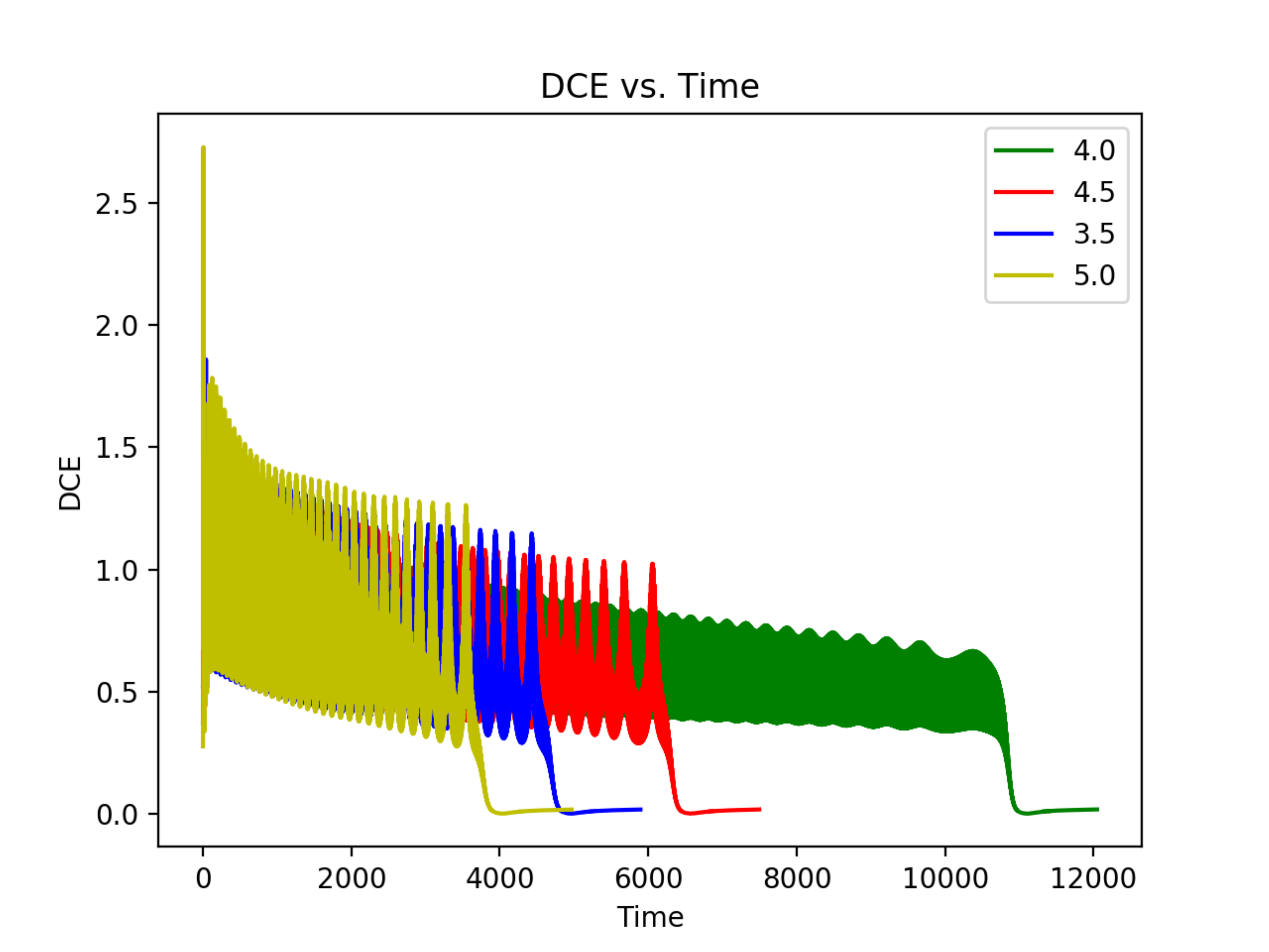}
\caption{DCE vs. time for several 3d ``normal'' oscillons with different initial radii $r_0$, as labeled.}
\label{3dCE}
\end{figure}

\begin{figure}[H]
   \centering
   \includegraphics[width=0.5\textwidth]{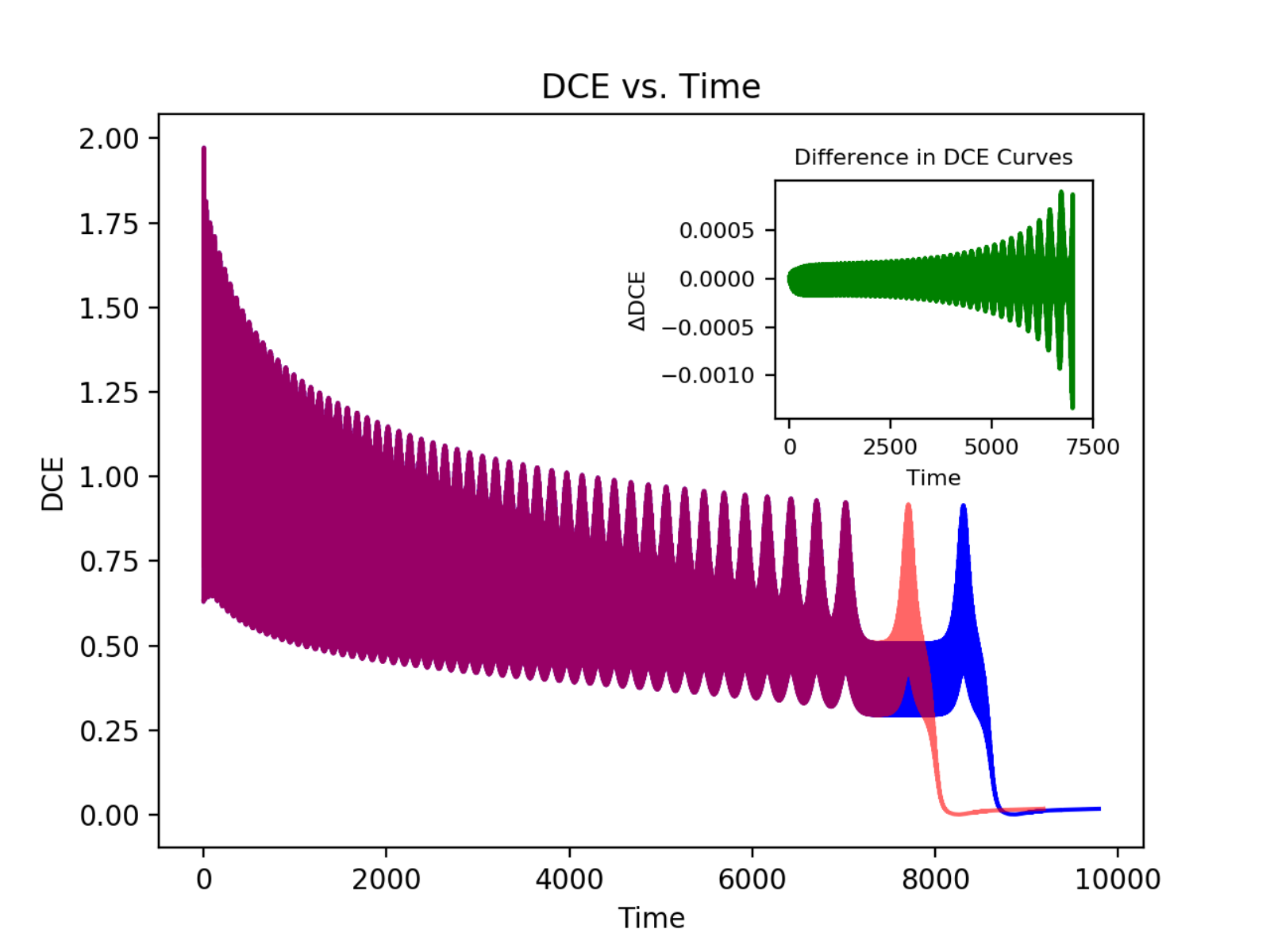}
\caption{DCE vs. time for two different 3d ``resonant oscillons'' lying on the 54th resonance.  The plateau regions late in the lifetime of the oscillons differ in length.  The inset shows that the difference between the two curve is small, but grows when approaching the plateau region.}
\label{DCERes54}
\end{figure}

 To get a better sense of resonant oscillons, we refer back to Figure \ref{resmountain}, the plot of oscillon lifetime as a function of initial bubble radius $r_0$, with spacing $\Delta r_0=0.00047$. In Ref. \cite{GleiserStephens} different measures derived from DCE were used to study the longevity of ``normal'' oscillons, that is, oscillons that don't live in one of the resonances anchored by a value $r_{0n}^*$. In Figures \ref{3dCE} and \ref{DCERes54}, we plot DCE vs. time for normal and resonant oscillons, respectively, for a few illustrative examples. In Figure \ref{3dCE}, we follow four normal oscillons with different initial radii: as seen for other physical systems and in Ref. \cite{GleiserStephens}, DCE is inversely correlated with stability, and, thus, to oscillon longevity. We will show that this remains true for resonant oscillons, although the situation is subtler, as can be seen in Figure \ref{DCERes54}: within our numerical precision, two resonant oscillons belonging to a specific resonance, $n = 54$, seem at least by eye to behave in essentially identical ways until the appearance of the last plateau, which extends further for the longer-lived oscillon, located higher up in the resonance peak. (See inset in Figure \ref{resmountain}.)

Figure \ref{DCERes54} and the inset show that by eye the behavior in DCE as a function of time for two oscillons on the same resonance seems nearly identical for most of their lifetimes.  We thus cannot use the measures of DCE from Ref. \cite{GleiserStephens} to study the longevity of resonant oscillons, since those relied on marked differences in the dynamics of normal oscillons throughout their lifetimes. (See Fig. \ref{3dCE}.)  However, if DCE is sensitive to growing instabilities in the configuration, it should be possible to construct a time-dependent measure that captures them.  In \cite{GleiserKrackow}, we found that an amplitude measure of the departure from virialization correlated well with resonant 3d oscillon longevity.  Following a similar logic, we constructed an average amplitude measure of the DCE, $\Delta \overline{C}$,  defined as \begin{equation}
\Delta \overline{C} = \frac{1}{T} \int_{t_{osc}}^{t_{fin}} \frac{\left| DCE(t) - \overline{C}(t) \right|}{\overline{C}(t)} dt.
\end{equation}
Here, $t_{osc}$ is chosen to be near the formation time of the oscillon.  We chose $t_{osc} = 1,000$ time units.  Testing $t_{osc}$ in the range $200 \leq t_{osc} \leq 1200$ showed that the results were consistent as long as $t_{osc} \gtrsim 800$ time units.  $t_{fin}$ is the time just before the oscillon decays, and $T$ is given by $T = \left| t_{fin} - t_{osc} \right|$.  $DCE(t)$ is the DCE as a function of time, and $\overline{C}(t)$ is the boxcar average of $DCE(t)$, where the boxcar interval is taken to be the period associated with the fast frequency of the DCE curve.  $\Delta \overline{C}$ is a normalized, time-averaged measure of the difference between DCE and its boxcar average during the oscillon's lifetime.  

\begin{figure}[H]
   \centering
   \includegraphics[width=0.5\textwidth]{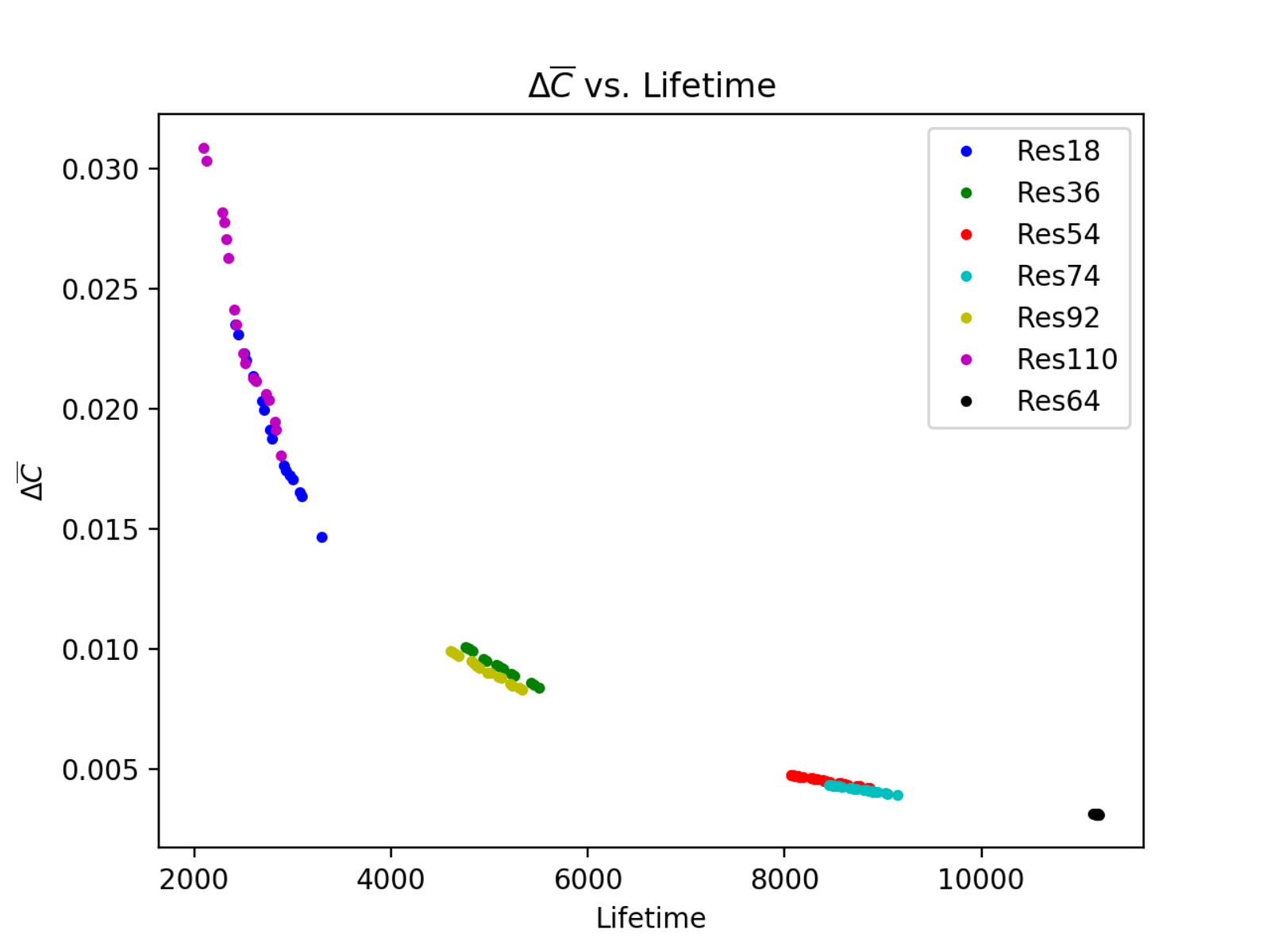}
\caption{$\Delta \overline{C}$ as a function of lifetime for several 3d resonant oscillons along various resonances, using $t_{osc} = 1000$.}
\label{C_Bar}
\end{figure}

As stated above, in \cite{GleiserStephens}, normal oscillon lifetimes were found to be inversely related to the magnitude of DCE measures.  Also, in \cite{GleiserKrackow} the departure from virialization was seen to be inversely related to lifetime.  Thus, we expect that $\Delta \overline{C}$ would also obey such a trend.  In Figure \ref{C_Bar} we plot $\Delta \overline{C}$ as a function of lifetime for oscillons along resonances 18, 36, 54, 74, 92, and 110.  Comparing with Figure \ref{resmountain}, we see that resonances higher up on the resonance mountain have lower values of $\Delta \overline{C}$, following a monotonic trend.  As discussed in Ref. \cite{GleiserKrackow}, the similar behavior of resonance pairings 18 - 110, 36 - 92, and 54 - 74 is due to the near symmetry of the resonance mountain at its peak and the fact that each of these resonance pairings is the same number of resonances away from the peak resonance on the resonance mountain - resonance 64.

\begin{figure}[H]
   \centering
   \includegraphics[width=0.5\textwidth]{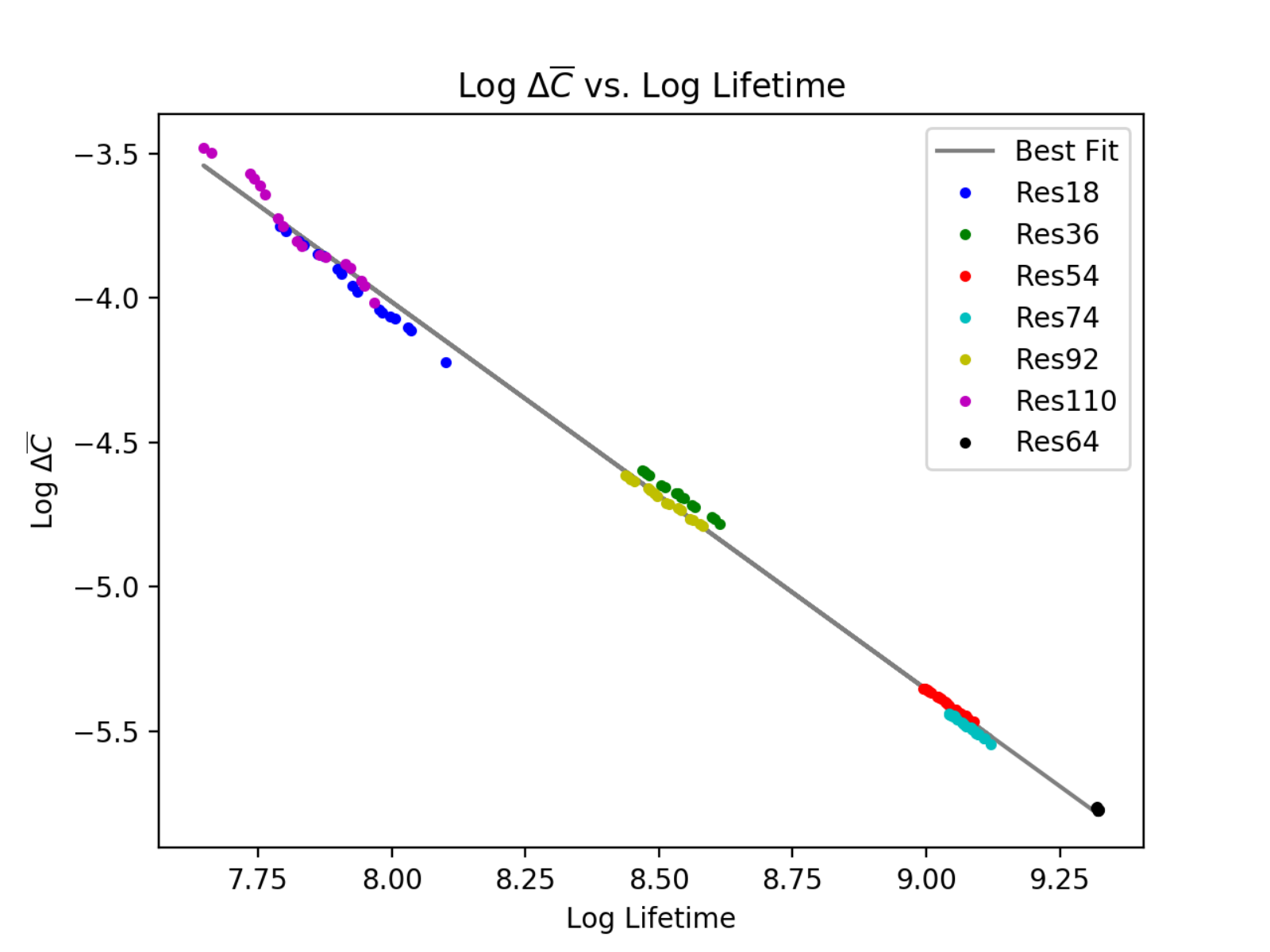}
\caption{Logarithmic plot showing a power law relating $\Delta \overline{C}$ with the lifetime of 3d resonant oscillons, $\Delta \overline{C} \sim b\tau^{\delta}$, where $\tau$ is the lifetime.}
\label{C_Bar_Log_Log}
\end{figure}

\par
Figure \ref{C_Bar_Log_Log} is the logarithmic plot of $\Delta \overline{C}$ as a function of lifetime for a few sample resonances.  The best fit is given by, 
\begin{equation}
\Delta \overline{C} = b\tau^{\delta},
\end{equation}
with $b=835.5$
and $\delta=-1.34$ as fitting constants. The coefficient of determination is $R^2=0.998$.

The power law relationship shows that as we explore oscillons higher in the resonance peaks, $\Delta \overline{C}$ continues to decrease as their lifetimes increase.  Extrapolation suggests, at least within our numerical accuracy, that oscillons located higher up in the resonances will have $\Delta \overline{C}\rightarrow 0$ as $\tau\rightarrow\infty$, corroborating the infinite lifetime conjecture.

\subsection{2d DCE Analysis}

\begin{figure}[H]
   \centering
   \includegraphics[width=0.5\textwidth]{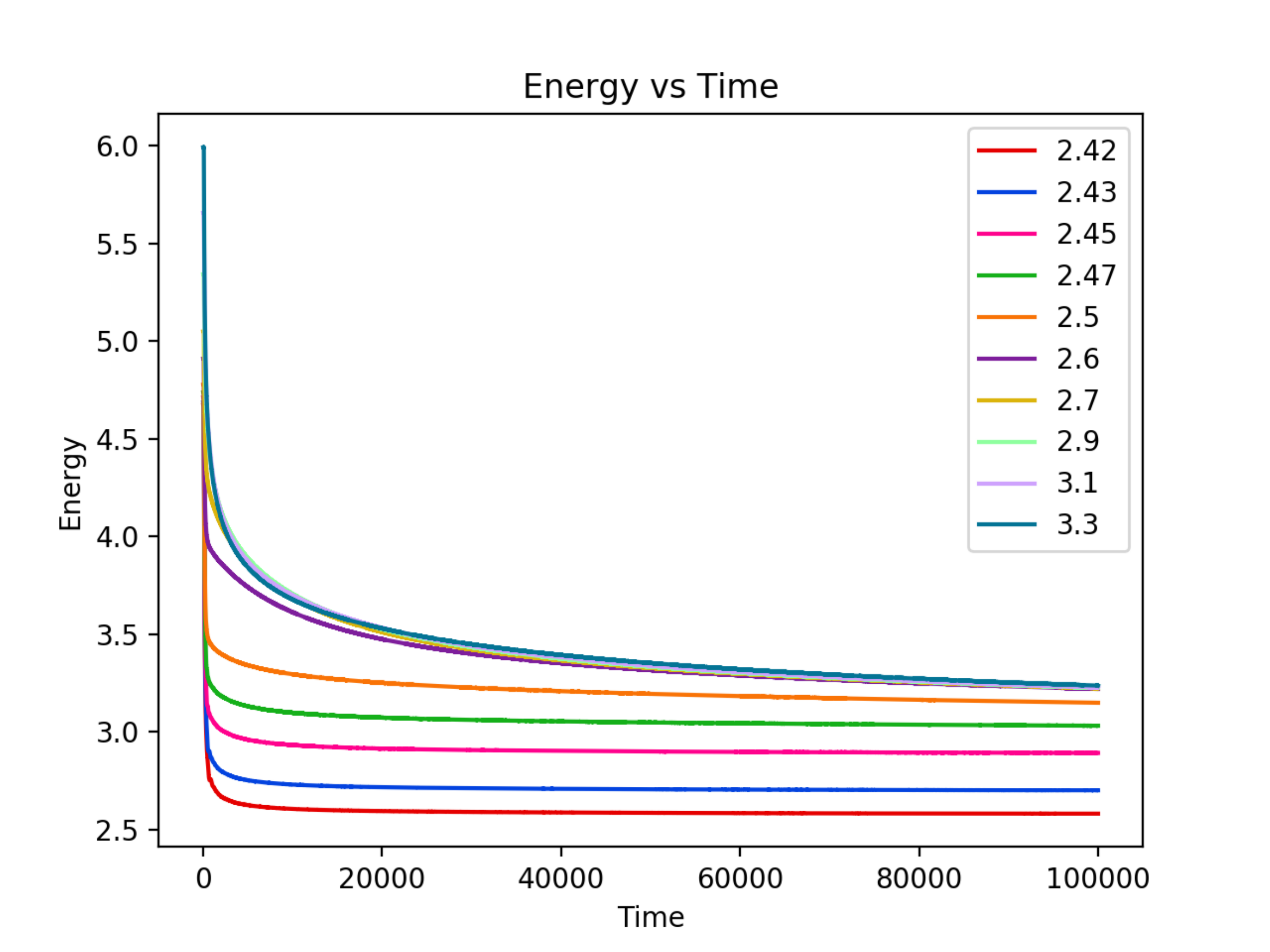}
\caption{Energy vs. time for 2d oscillons with different initial radii.}
\label{2dEnergy}
\end{figure}

Multiple numerical studies investigating oscillons in 2d found that out to $10^7$ time units, oscillons do not decay \cite{Sornborger, Salmi}.  Here, we will advance this study exploring the fact that although 2d oscillons may be infinitely long-lived and, therefore, cannot be distinguished from one another by their lifetimes, they do not all share the same plateau energy, as do oscillons in 3d.  This can be seen in Figure \ref{2dEnergy}, where energy is plotted out to $10^5$ time units for several oscillons with different initial radii.  To study the longevity of 2d oscillons, we must investigate their properties as a function of their distinct plateau energies, in particular, whether they radiate towards an attractor solution, as conjectured analytically in Ref. \cite{Sicilia}.

In Ref. \cite{GleiserKrackow}, we found a correspondence between 2d oscillons' plateau energies and their stabilities.  Upon performing a parametric resonance analysis, we saw that the Floquet exponent of the solution to the linearized equations grew with increasing plateau energy, concluding that 2d oscillons with higher plateau energies are less stable.  Above, in 3d, we found that longer-lived, and thus more stable, oscillons exhibited smaller values of $\Delta \overline{C}$.  Since our previous findings suggest that both plateau energy and DCE can be used to study the stability of resonant oscillons, we expect that 2d oscillons with lower plateau energies will also exhibit smaller values of different DCE measures.

To implement this analysis, we chose the plateau energy to be the value of the oscillon energy at 100,000 time units, and defined a new measure of DCE, called ${\rm DCE_{\rm Min}}$.  This is the minimum value of DCE that an oscillon obtains between 20,000 and 100,000 time units.  We choose 20,000 time units as a lower bound in time when searching for ${\rm DCE_{\rm Min}}$ to allow the oscillon to shed most of its initial energy, as can be seen in Figure \ref{2dEnergy}.  After 20,000 time units, radiation rates decrease rapidly (but not completely) for all oscillons.

\begin{figure}[H]
   \centering
   \includegraphics[width=0.5\textwidth]{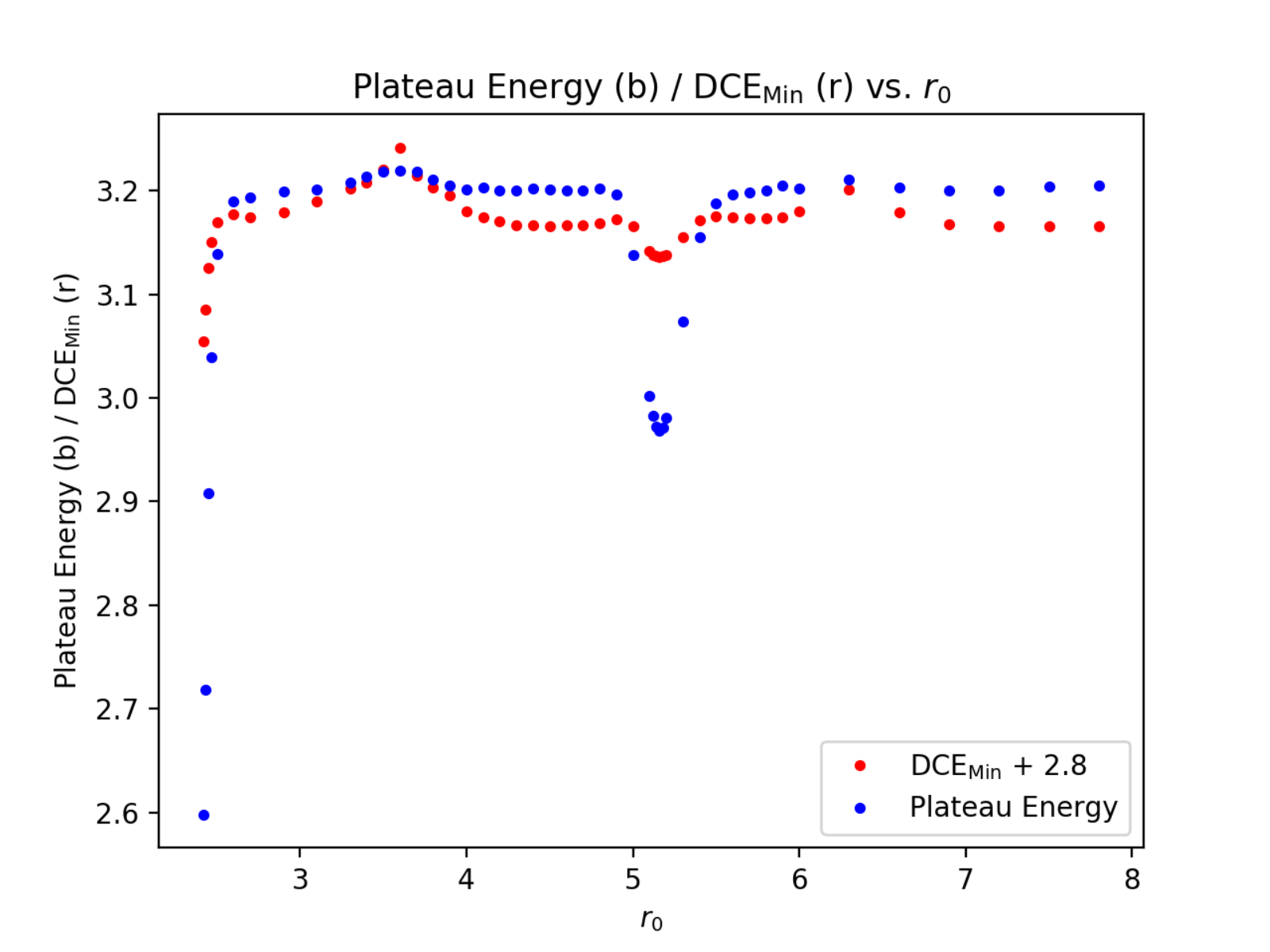}
\caption{Plateau energy (blue) and ${\rm DCE_{\rm Min}}$ (red) as a function of initial radius $r_0$.  ${\rm DCE_{\rm Min}}$ values are vertically displaced upwards by 2.8 units to display similar trends of ${\rm DCE_{\rm Min}}$ and plateau energy.}
\label{DCE_and_Plat_E_vs_R_0}
\end{figure}

In Figure \ref{DCE_and_Plat_E_vs_R_0} we display the plateau energy (blue) and  ${\rm DCE_{\rm Min}}$ (red) as a function of $r_0$.   ${\rm DCE_{\rm Min}}$ is vertically displaced 2.8 units upwards so that similarities in the trends between  ${\rm DCE_{\rm Min}}$ and plateau energy can be observed.   The trend is quite clear, as  ${\rm DCE_{\rm Min}}$ qualitatively tracks plateau energies.  Also in this plot, we see that there are many oscillons with plateau energies $E_{\rm plat} \sim 3.2$.  (We are currently investigating the curious dip around $r_0\sim 5.2$.) From Figure \ref{2dEnergy}, we can see that the oscillons with initial radii $r_0 = 2.6, \; 2.7, \; 2.9, \; 3.1,\textrm{ and } 3.3$ are an illustrative sample of the many that group around $E_{\rm plat} \sim 3.2$.  The corresponding values of  ${\rm DCE_{\rm Min}}$ hover slightly below $\sim 0.38$.  

\begin{figure}[H]
   \centering
   \includegraphics[width=0.5\textwidth]{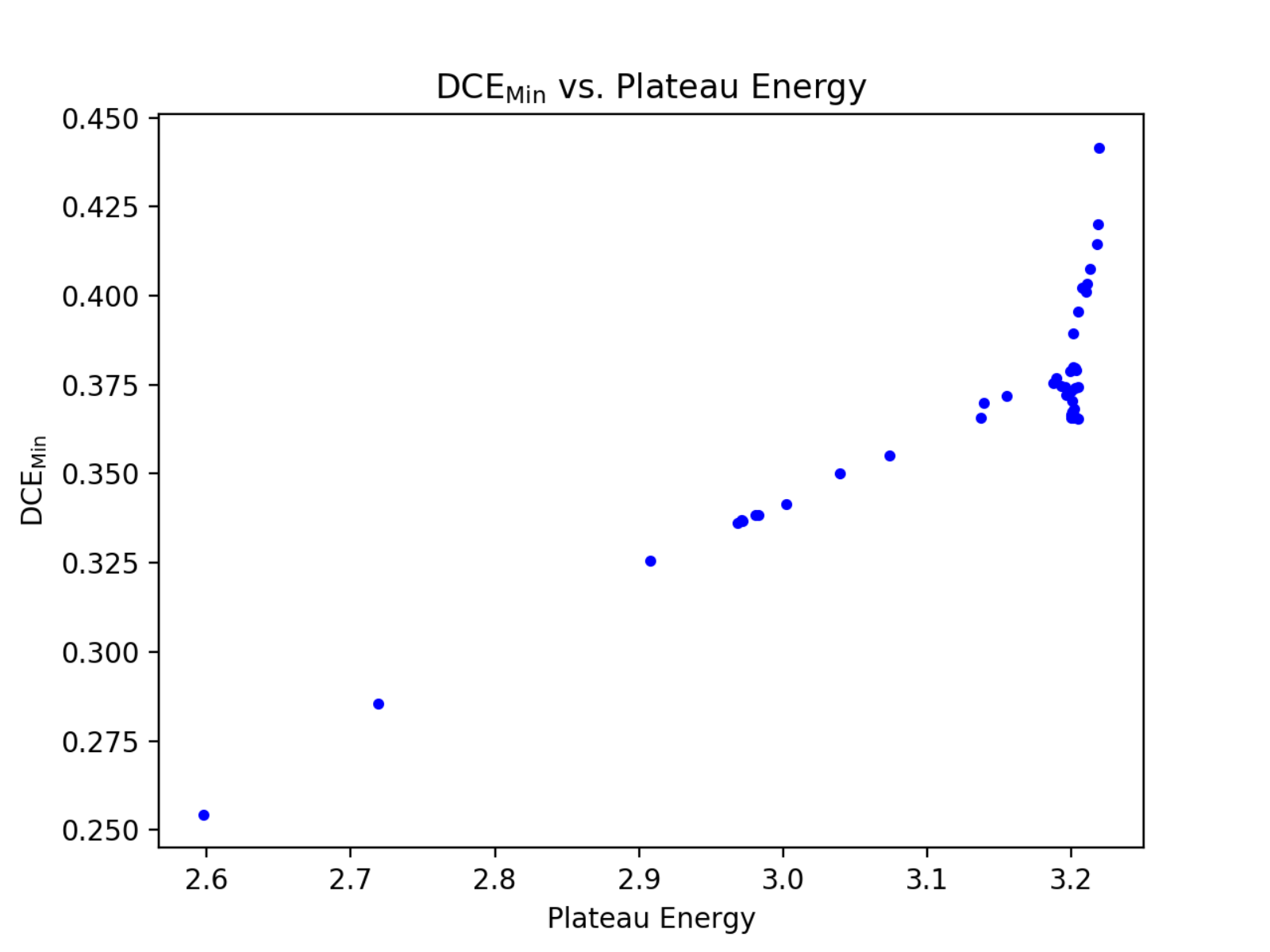}
\caption{ ${\rm DCE_{\rm Min}}$ vs. Plateau Energy for 2d oscillons.}
\label{DCE_Min_vs_E_Plat}
\end{figure}

Figure \ref{DCE_Min_vs_E_Plat} displays  ${\rm DCE_{\rm Min}}$ as a function of plateau energy for the same 2d oscillons displayed in Figure \ref{DCE_and_Plat_E_vs_R_0}.  As expected from Figure \ref{DCE_and_Plat_E_vs_R_0}, as  ${\rm DCE_{\rm Min}}$ increases so does the plateau energy.  However, note the cluster of oscillons around a plateau energy of $\sim 3.2$ and  ${\rm DCE_{\rm Min}}\sim 0.375$.  The cluster divides two different trends, characterized by approximately linear slopes. To understand this behavior, we investigated the rates of energy loss for the different oscillons.  

\begin{figure}[H]
   \includegraphics[width=0.5\textwidth]{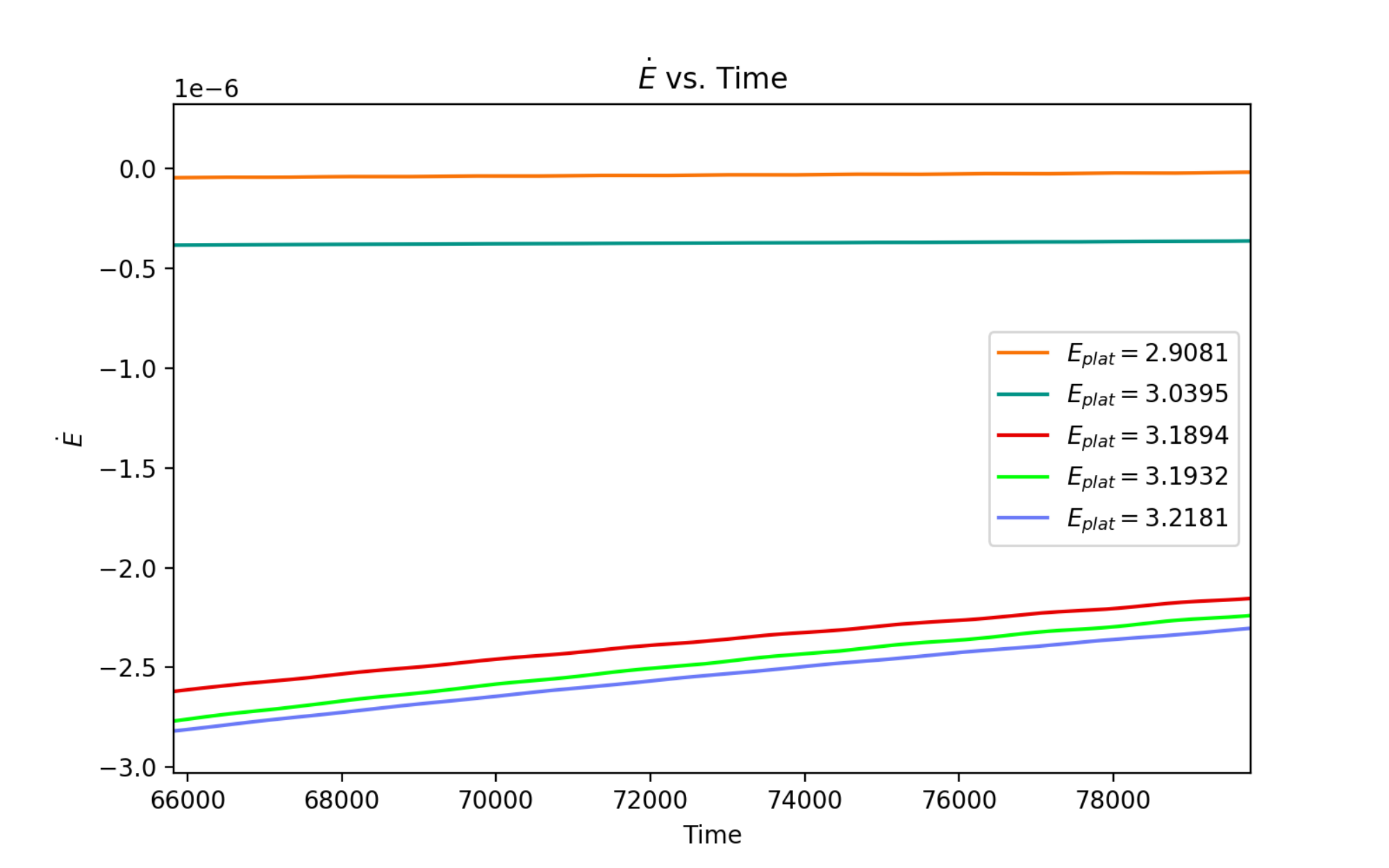}
   \caption{Oscillon radiation rate, $\dot{E}$, as a function of time for several 2d oscillons.}
\label{E_dot}
\end{figure}

Figure \ref{E_dot} shows the rate of energy loss $\frac{dE}{dt} \equiv \dot{E}$ for five sample oscillons as a function of time.  The oscillon with plateau energy, $E_{\rm plat} = 3.1894$ (red curve) is found in the cluster region in Figure \ref{DCE_Min_vs_E_Plat}.  Back to Figure \ref{E_dot}, oscillons with plateau energies lower than $E_{\rm plat} = 3.1894$ have slower radiation rates (top of figure), while oscillons with plateau energies above $E_{\rm plat} = 3.1894$ have faster radiation rates (bottom of figure).

In Figure \ref{DCE_Min_vs_E_Plat}, the slow (fast) increase in ${\rm DCE_{\rm Min}}$ before (after) the cluster can be attributed to different radiation rates.  Relative to the radiation rates for oscillons within the cluster, those for oscilons below the cluster are monotonically slower, while those above the cluster are monotonically faster.  We also note that the radiation rates decrease with time in all cases. Our results indicate that increasing values of the plateau energy are correlated with increasing ${\rm DCE_{\rm Min}}$ and with increasing radiation rates, indicating growing instability. This correlation between plateau energy, configurational entropy, and radiation rate helps us define the meaning of instability in the context of 2d oscillons. Comparing with Figure \ref{2dEnergy}, we see that oscillons with lower plateau energies settle to those near constant plateau energies very quickly and then hardly radiate, whereas oscillons with higher plateau energies shed away energy faster.  In Figure 9, notice that the oscillon with plateau energy, $E_{\rm plat} = 2.9081$ radiates very slowly.  Extrapolation suggests that the energies converge to an asymptotic value. Indeed, in Ref. \cite{Sicilia}, it was argued that 2d oscillons may be infinitely long-lived because they asymptotically approach an attractor solution with energy $E_{\rm attract}$. Using the formalism of Ref. \cite{Sicilia} with the potential of Eq. 4, we find $E_{\rm attract} = 2.2229$.  From Figure \ref{2dEnergy} we see that the lowest oscillon we probed, with $r_0=2.42$, has plateau energy $E_{\rm plat} \simeq 2.5978$, and thus still larger than $E_{\rm attract}$. The closer to the attractor solution, the slower oscillons radiate, as seen in
Figure \ref{E_dot}.

\section{Summary and Conclusions}

We investigated the longevity of two classes of oscillons: resonant oscillons in three spatial dimensions and oscillons in two dimensions using a measure of spatial complexity known as configurational entropy (CE) \cite{GleiserStamatopoulos1}. Resonant oscillons are configurations that live on specific regularly-spaced resonant peaks in the lifetime versus initial radius plot of oscillons (see Figure \ref{resmountain}) that have been conjectured in Ref. \cite{Honda} to have potentially infinite lifetimes. We found a power law relating a stability measure derived from CE called $\Delta {\bar C}$ and the oscillons' lifetimes, shown in Figure \ref{C_Bar_Log_Log}. Upon extrapolation and within the limits of our numerical approach, the trend does support the conjecture, in agreement with our previous results using dynamical methods \cite{GleiserKrackow}. In two dimensions, oscillons have not been seen to decay in numerical studies. We found a correlation between the oscillons' plateau energies, a CE-derived measure called ${\rm DCE}_{\rm Min}$, and their radiation rates that supports the conjecture of Ref. \cite{Sicilia} that 2d oscillons tend to a classically stable attractor solution.

\end{document}